\documentclass[12pt]{iopart}
\usepackage{times}
\usepackage{graphicx}

\newtheorem{example}{Example}
\newtheorem{dex}{Definition}
\newtheorem{theorem}{Theorem}
\def\ket#1{| #1 \rangle}
\begin{document}
\title{Entanglement Dissipation: Unitary and Non-unitary Processes}
\author{Allan I. Solomon}
\address{Department of Physics and Astronomy, The Open University, MK7 6AA, UK \\and \\LPTMC, Universit\'e  de Paris VI, France}
\begin{abstract}
Dissipative processes in physics are usually associated with non-unitary actions.  However, the important resource of entanglement is not invariant  under general unitary transformations, and is thus  susceptible to unitary ``dissipation''.  In this note we discuss both unitary and non-unitary dissipative processes, showing that the former is ultimately of value, since reversible, and enables the production of entanglement; while even in the presence of the latter, more conventional non-unitary and non-reversible, process there exist  nonetheless invariant entangled states.
\end{abstract}
\section{Introduction to Bipartite Entanglement}
\subsection{Definition and Measure (Concurrence)}
Bipartite entanglement involves the direct product space $V$ of two (complex) vector spaces, $V_1$ and $V_2$, of dimension $m$ and $n$ respectively.
If $V_1$ has basis $\{e_i,\; i=1\ldots m\}$ and $V_2$  has basis $\{f_j,\; j=1, \ldots n \}$  then $\{e_i \otimes f_j, \;\;\; i=1\ldots m,j=1 \ldots n\}$ is a basis for $V = V_1\otimes V_2$.

In this note we specialize to the specific and more familiar case of two two-qubit spaces, $m=n=2$, with the same standard bases $\{ e_i (=f_i), i=1,2\}$ for $V_1$ and $V_2$. We shall also use the matrix forms $\{e_1,e_2\}=\{[1,0],[0,1]\}$ \footnote{For typographical simplicity we write all our (column) vectors as row vectors.} as well as the ket notation $e_1=|0\rangle$ and $e_1 \otimes e_1 = |0,0\rangle$ etc.  Note that in the context of quantum mechanics we refer to vectors as {\em pure} states.
\begin{dex}[Entangled pure state]\label{entangled}
Every vector in $V_1\otimes V_2$ is a sum of products; but not every vector is a product.  If it is a product, then it is said to be {\em non-entangled} or {\em separable}.
\end{dex}
It is a straightforward matter  to determine whether a vector $v \in V$ is entangled or not.
\newline \newline
If $v \in V$ is non-entangled, i.e. separable, then
\begin{eqnarray}
  v \in V_1\otimes V_2 &=& \sum_{i,j=1}^{2}c_{ij} e_i \otimes e_j\;\;\;\; \nonumber \\
   &=& (x_{1}e_{1}+x_{2}e_{2})\otimes(y_{1}e_{1}+y_{2}e_{2}) \nonumber\\
  \Rightarrow &c_{ij}&=x_{i}y_{j} \;\;\;\; \{i,j=1,2\}
\end{eqnarray}
from which we deduce that the matrix $c$ of coefficients $c_{ij}$ has determinant zero, $\det c=0$.
\begin{example}[Separable pure state]
Consider the bipartite pure state
\begin{equation}\label{ex1}
    (1/\sqrt{50})(|0,0\rangle+3|0,1\rangle+2|1,0\rangle+6|1,1\rangle).
    \nonumber
\end{equation}
The matrix $c$ of coefficients is given by
\begin{equation}\label{C1}
    c=1/\sqrt{50}\left[
    \begin{array}{cc}
     1& 3 \\
     2&6
     \end{array}
     \right]
     \nonumber
\end{equation}
for which $\det c=0$ and so the state is separable.
\end{example}
\begin{example}[Bell state]\label{Bell}
An example of a maximally entangled two-qubit state is given by the Bell state
$\frac{1}{\sqrt{2}}(|0,0\rangle+|1,1 \rangle)$
for which
\begin{equation}\label{C2}
    c=\frac{1}{\sqrt{2}}\left[
    \begin{array}{cc}
     1& 0 \\
     0&1
     \end{array}
     \right]
     \nonumber
\end{equation}
and so $\det c=1/2$.
\end{example}
This simple criterion  for pure state separability in fact gives a {\em measure} of entanglement for pure states. To obtain this measure, we normalize so that a Bell state, such as that of  Example \ref{Bell}, has maximal  measure of entanglement equal to  $1$, and we arrive at the definition of the {\em concurrence} applicable to pure states:
\begin{dex}[Pure state Concurrence]\label{concurrence}
A measure of entanglement for pure bipartite states (belonging to two two-qubit spaces) is given by the {\em concurrence} $\mathcal{C}=2|\det c|.$
\end{dex}

The concurrence $\mathcal{C}$ is essentially equivalent to the measure $\mathcal{E}$ called {\em Entanglement of Formation}, based on the Von Neumann entropy of the partial trace\cite{nc}\footnote{Writing $f=-x log_{2}x-(1-x) log_{2}(1-x)$ then $\mathcal{E}=f((1+\sqrt{1-{\mathcal{C}}^{2}})/2)$.}.

The entanglement measure $\mathcal{C}$ may be extended to {\em general}, or mixed, states (density matrices). We describe this extension in Section \ref{mixcon}. 
\subsection{General states}
We are initially concerned in this note with {\em pure} states; i.e. represented by vectors $v$ in $V$.  However, we can equally represent our vector $v$ by the  matrix $\rho(v) =v v^{\dagger}$. {Of course the overall irrelevant phase information is lost in this form.}  It is easily verified that $\rho$ is a hermitian matrix of rank one, that is, all sub-matrices of order $2$ or more  have determinant zero.  And it has a sole non-zero eigenvalue which is equal to $1$. It has trace equal to one, assuming that $v$ is normalized.  A hermitian matrix all of
whose eigenvalues are $\geq 0$ is called a {\em positive matrix} (more accurately, semi-positive). We may extend this description of the matrix associated with a pure state to give the following definition of a {\em state} in general ({\em mixed state} or {\em density matrix}):
\begin{dex}[State]\label{state}
  A state $\rho$ (acting on a space $V$) is a positive matrix of trace 1.
\end{dex}
Equivalently,
\begin{dex}[State as convex sum]
  A state $\rho$ is a convex sum $\sum_{i}\lambda_i \rho_i\; \;  (\lambda_i \geq 0\; \; \; \sum_{i}\lambda_{i}=1)$ of pure states $\rho_i$.
\end{dex}
We simply note here the definition of separability for (general) states:
\begin{dex}[Separable state]\label{sep}
  The  state $\rho$ acting on $V_1 \otimes V_2$ is said to be {\em separable} if is given by a convex sum $\sum_{i}\lambda_i \rho^{1}_i \otimes \rho^{2}_i\; \;  (\lambda_i \geq 0\; \; \; \sum_{i}\lambda_{i}=1)$ where $\rho^{\alpha}_i$ acts on $ V_\alpha$ .
  \end{dex}
  When $\rho=\rho^{1} \otimes \rho^{2}$ it is said to be {\em simply separable}.
  The above definition extends immediately to {\em multipartite} states.

  If we have a measure of entanglement $\mathcal{E}$ for pure states (such as that given in Definition~\ref{concurrence}) we may extend it to general states by
  \begin{dex}[Entanglement of general state]\label{entang}
   The entanglement ${\mathcal E}(\rho)$ of a mixed bipartite state   $\rho$ acting on $V_1 \otimes V_2$ is given by $ {\cal E} (\rho)= \min\{\sum_i \mu_i {\cal E}(\psi_i) | \rho= \sum_i \mu_i \rho(\psi_i)\}$ where the $\psi_i$ are pure states in $V_1 \otimes V_2$.
\end{dex}
\subsection{Unitary and Local Unitary Transformations}
Since every (normed) vector $v \in V$ can be transformed to the (non-entangled) state $|0,0\rangle$ by a unitary transformation, it is clear that entanglement is {\em not} invariant under unitary transformations.  However, under a local unitary transformation, defined by $U = U_1 \otimes U_1$, one can see that the concurrence, for example, is invariant:
\begin{theorem}\label{conc}
The concurrence $\mathcal{C}$ is invariant under local unitary transformations. \newline
Let $v=\sum_{i,j=1\ldots 2} a_{ij}e_i \otimes e_j \in V=V_1 \otimes V_2$, and the unitary matrix $U=U_1 \otimes U_2$ be a local unitary matrix; then
\begin{eqnarray*}
  Uv &=& \sum a_{ij} U_1 e_i \otimes U_2 e_j\\
    &=& \sum a_{ij} (U_1)_{ik} e_k \otimes (U_2)_{jr} e_r \\
   &=& \sum c_{kr}e_k \otimes e_r
\end{eqnarray*}where  $c_{kr}=\sum_{ij}{a_{ij} (U_1)_{ik}(U_2)_{jr}}$
so that $c = \tilde{U_1} a U_2$
whence
\begin{eqnarray*}
   |\det c | &=& |\det(\tilde{U_1} a U_2)| \\
             &=& |\det{\tilde{U_1}}|\;|\det a|\;|\det{U_2}| \\
             &=&  |\det a| \; \; \;  {\rm since \; \; \; } |\det{U_i}|=1.
\end{eqnarray*}
\end{theorem}

We may see rather more immediately from Definition \ref{sep} that the property of {\em being} separable is invariant under local unitary transformations; and this extends to the multipartite case. However, an extension of Theorem \ref{conc} to multipartite systems, namely that such local transformations preserve the {\em measure} of entanglement, would depend on a definition of measure (or measures) of entanglement for such systems, which is currently unavailable. For general multipartite states, local unitary equivalence does not preserve all the relevant (state and substate) entanglement properties \cite{hso}.

\section{Unitary Dissipation}
Although the notion of dissipation  is more usually associated with a non-unitary process, from the preceding we see that {\em entanglement} is subject to unitary dissipation, since unitary evolution associated with a (hermitian) hamiltonian does not necessarily preserve entanglement.  Of course, the good news is the other side of this coin; that is, entanglement may be produced by the evolution induced by  a quantum control hamiltonian.  Quantum control applied to multipartite systems has been well treated, see for example \cite{ss}.  We choose a simple example to illustrate the  bipartite case.
\begin{example}[Entanglement production and decay]
Consider the unitary evolution $U(t)$ induced by the  hamiltonian $H$ given by
\begin{equation}\label{ham}
   H=  \left[ \begin {array}{cccc} {\it x_1}&0&0&y\\\noalign{\medskip}0&{\it
x_2}&0&0\\\noalign{\medskip}0&0&{\it x_3}&0\\\noalign{\medskip}y&0&0&{\it x_1}\end {array} \right].
\end{equation}
 This is essentially a free hamiltonian with the addition of an off-diagonal time-independent control term $y$\footnote{For calculational simplicity we have chosen a degeneracy between the first and last energy levels, $x_1=x_4$.}.  Note that without this latter term $H$ would not change the entanglement since $H$ would then be a {\em local transformation}.

 We act by the unitary evolution matrix $U(t) = \exp(itH)$ induced by Eq.(\ref{ham}) on the base vector $v_0\equiv [1,0,0,0]$. (Note that for such calculations it is important to choose a fixed basis - here the standard basis.)
\begin{eqnarray}
 v(t)&=&\exp{(itH)}v_0 \\
   &=&e^{it {\it x_1}}\, [\cos \left( t\,y \right) ,0,0,
i\sin \left( t\,y \right) ].
\end{eqnarray}
 Apart from an overall phase factor, only the control term $y$ plays a r\^{o}le in the entanglement production.

Using the measure of entanglement  for pure states given in Definition \ref{concurrence},  the concurrence  for $v(t)$ is given by $|\sin(2ty)|$ (See Figure 1).
\end{example}
\subsection{Unitary dissipation of Entanglement}
\begin{example}[Unitary dissipation of entanglement]
Referring to the previous example and Figure 1, we see immediately that at $t=\pi/4$ (in units of $1/y$ where                    $y$ is the control frequency) we have the maximally entangled (Bell) state $ \frac{1}{\sqrt{2}}[ 1,0,0,i]$ (up to an overall phase factor). The unitary action $U(t)$ destroys the entanglement, completely at $t=\pi/2$.
\end{example}
\begin{figure}
\vspace{1cm}
\begin{center}\resizebox{8 cm}{!}
{\includegraphics{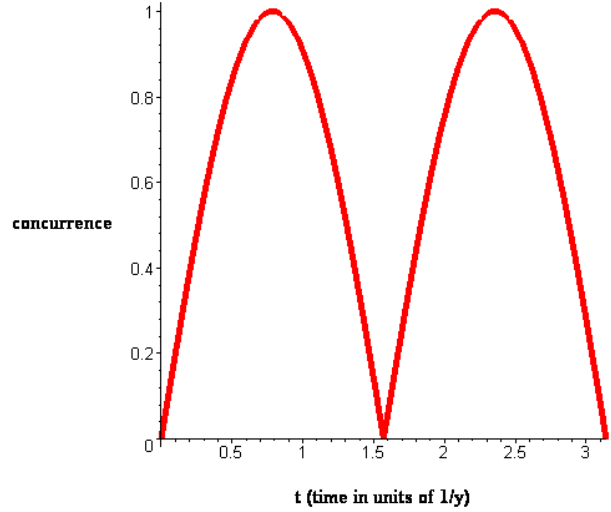}}
\caption{Concurrence {\it v} Time t (units of 1/control frequency)}
\end{center}
\label{conplot}
\end{figure}
\section{Environmental Dissipation}\label{envir}
For the usual description of dissipative processes we must use the standard definition of a
{\em general} quantum state $\rho$ given by Definition \ref{state}.  Thus $\rho$ is an $N \times N$ positive matrix  (and for our two-qubit examples, $N=4$).
For a non-dissipative process, the basic equation which determines the evolution of a hamiltonian
quantum  system  may be written in the form of a differential equation for the quantum
state, the Liouville-Von Neumann equation\cite{bp} (choosing units in which $\hbar=1$):
\begin{equation} \label{lse}
  i \frac{d}{dt}{\rho}(t) = [{H},{\rho}(t)]
  \equiv {H}{\rho}(t) - {\rho}(t){H}
\end{equation}
where $H$ is the total hamiltonian of the system.
The standard form of a general {\em dissipative} process in Quantum Mechanics is governed by
the Liouville equation  obtained by adding a dissipation (super-)operator
${L}_D[{\rho}(t)]$ to Eq.(\ref{lse}):
\begin{equation} \label{eq:dLE}
   i\dot{\rho}(t) = [{H},{\rho}(t)] + i {L}_D[{\rho}(t)].
\end{equation}
\subsection{Liouville Dissipation}
In general, uncontrollable interactions of the system with its
environment lead to two types of dissipation: phase decoherence
(dephasing) and population relaxation.

Phase decoherence occurs when the
interaction with the environment destroys the phase correlations
between states, which leads to changes in the off-diagonal elements
of the density matrix:
\begin{equation} \label{eq:dephasing}
 \dot{\rho}_{kn}(t)
  = -i([{H},{\rho}(t)])_{kn}-\Gamma_{kn}\rho_{kn}(t)
\end{equation}
where $\Gamma_{kn}$ (for $k\neq n$) is the dephasing rate between
$\ket{k}$ and $\ket{n}$.

Population relaxation occurs, for instance, when a
quantum particle in state $\ket{n}$ spontaneously emits a photon and
moves to another quantum state $\ket{k}$, which changes the
populations according to
\begin{equation} \label{eq:poptrans}
\dot{\rho}_{nn}(t) = -i([{H},{\rho}(t)])_{nn}
  +\sum_{k\neq n} \left[\gamma_{nk}\rho_{kk}(t)-\gamma_{kn}\rho_{nn}(t)\right]
\end{equation}
where $\gamma_{kn}\rho_{nn}$ is the population loss for level
$\ket{n}$ due to transitions $\ket{n}\rightarrow\ket{k}$, and
$\gamma_{nk}\rho_{kk}$ is the population gain caused by transitions
$\ket{k}\rightarrow\ket{n}$.  The population relaxation rate
$\gamma_{kn}$ is determined by the lifetime of the state $\ket{n}$,
and for multiple decay pathways, the relative probability for the
transition $\ket{n}\rightarrow\ket{k}$.  

Phase decoherence and
population relaxation lead to a dissipation super-operator
(represented by an $N^2 \times N^2$ matrix) whose non-zero elements
are
\begin{equation}\label{supop}
  \begin{array}{ll}
  ({L}_D)_{[k;n],[k;n]} = -\Gamma_{kn} & k \neq n \\
  ({L}_D)_{[n;n],[k;k]} = +\gamma_{nk} & k \neq n \\
  ({ L}_D)_{[n;n],[n;n]} = - \sum_{n\neq k} \gamma_{kn}
 \end{array}
\end{equation}
where $\Gamma_{kn}$ and $\gamma_{kn}$ are taken to be positive numbers, with
$\Gamma_{kn}$ symmetric in its indices.

In Eq.(\ref{supop} we have introduced the
convenient notation $[m;n] = (m-1)N+n$.
The $N^2 \times N^2$ matrix super-operator $L_D$ may be thought of as
acting on the $N^2$-vector ${\bf r}$ obtained from $\rho$ by
\begin{equation}\label{vec}
  {\bf r}_{[m;n]} \equiv \rho_{mn}.
\end{equation}
The resulting vector equation is
\begin{equation}\label{veceq}
  \dot{\bf{r}} = L {\bf{r}} = (L_H+L_D) {\bf{r}}
\end{equation}
where $L_H$ is the anti-hermitian matrix corresponding to the
hamiltonian $H$.
We obtain $L_H$ explicitly by using  the standard algebraic trick applied in evaluating
Liouville equations (see, for example~\cite{havel}).  The
correspondence between $\rho$ and $\bf {r}$ as given in
Eq.~(\ref{vec}) tells us, after some  manipulation of indices, that
\begin{equation}\label{trick}
   \rho \rightarrow {\bf {r}} \Rightarrow A \rho B \rightarrow A \otimes \tilde{B}\; \;  {\bf {r}}
\end{equation}
using the direct (Kronecker) product of matrices.
\section{Physical Processes}\label{phys}
The quantum Liouville equation (\ref{eq:dLE}) is very formal; it covers both physical and non-physical processes  and  may tell us  little about an actual physical dissipation process.  For example, the {\em values} of the dissipation parameters $\Gamma_{kn}$ and $\gamma_{kn}$ are not determined and a general choice will not lead to a physical process - that is, one under which the state $\rho(t)$ remains a physical state - unless the parameters satisfy various constraints\cite{constraints}.  This  (completely) positive evolution and the appropriate constraints  emerge from physical stochastic dissipation equations such as those given by Lindblad and others in differential form \cite{Lindblad}, as well as in global form \cite{Kraus}.  Nevertheless, the  virtue of Eq.(\ref{eq:dLE}) is that essentially every dissipation process will have to satisfy it and so results derived from its use will have great generality.

Since in our examples we wish to restrict ourselves to {\em physical} processes, we obtain our dissipation parameters $\Gamma$ and $\gamma$ by use of the Lindblad equation.
\subsection{Lindblad Equation}
Completely positive evolution of the system is guaranteed by the Lindblad form of the
dissipation super-operator $L_D$
 \begin{equation}     \label{lin}
  L_D[{\rho}(t)] =
  \frac{1}{2} \sum_{s=1}^{N^2} \left\{ [{V}_s{\rho}(t),{V}_s^\dagger] +
                               [{V}_s,{\rho}(t){V}_s^\dagger] \right\}
\end{equation}
where the matrices $V_s$ are arbitrary. The standard basis for $N \times N$ matrices is
given by
\begin{equation}\label{sb}
 (E_{ij})_{mn}={\delta}_{im}{\delta}_{jn}   \;\;\;\;\;\;(i,j,m,n=1 \ldots N)
\end{equation}
Relabelling, using the notation $[m;n] = (m-1)N+n$, we choose
\begin{equation}\label{av}
  V_{[i;j]}=a_{[i;j]}E_{[i;j]} .
\end{equation}
All the $\Gamma$'s and $\gamma$'s are determined by the (absolute values of) the $N^2$ (=16 here)  parameters $a_{[i;j]}.$
\subsection{Pure decoherence only}
So far we have discussed the most general case, when in principle all relaxation and
decoherence parameters may be present in the dissipation matrix. However, experimentally, the relaxation time $T_1$ for most
systems is  much longer than the dephasing time $T_2$ so that we may effectively neglect
the relaxation rates $\gamma$.
In the {\em pure decoherence (dephasing)} case, comparison of Eq.(\ref{lin}) and Eq.(\ref{av})  with Eq.(\ref{supop}) tells us that the $\gamma$ terms vanish if we choose $a_{[i;j]}=0$ for $i\neq j$.
The decoherence parameters $\Gamma_{ij}$ are then given by
\begin{equation}\label{gammas}
 \Gamma_{ij} = \frac{1}{2}(|a_{[i;i]}|^2 + |a_{[j;j]}|^2)\; \; \; \; (i,j=1 \ldots N\;\;\;i \neq j).
\end{equation}

This leads to a mathematically very simple situation, as
the dissipation matrix $L_{D0}$ is then diagonal.   For the $N=4$ system, this gives 6 pure dephasing parameters ($\Gamma_{ij}=\Gamma_{ji}$ and $\Gamma_{ii}=0$), determined by 4 constants, so there are two relations between the $\Gamma$'s - see Eq.(\ref{Gs}) below.

 Explicitly for the two-qubit, 4-level case,
\begin{eqnarray}\label{LD0}
  L_{D0}&=&{\rm diag} \{0,-\Gamma_{{12}},-\Gamma_{{13}},-\Gamma_{{14}},-\Gamma_{{21}},0,-\Gamma_{{23}},-\Gamma_{{24}}, \nonumber \\
        & &-\Gamma_{{31}},-\Gamma_{{32}},0,-\Gamma_{{34}},-\Gamma_{{41}},-\Gamma_{{42}},-\Gamma_{{43}},0
\}
\end{eqnarray}
with  $\Gamma_{ij}=\Gamma_{ji}$.

The constraints imposed by the physical process are
\begin{equation}\label{Gs}
   \Gamma_{12}+\Gamma_{34}=\Gamma_{14}+\Gamma_{23}=\Gamma_{13}+\Gamma_{24}.
\end{equation}
\subsection{Concurrence}\label{mixcon}
In Sections \ref{envir} and \ref{phys} we are perforce dealing with general states, and so we must use the extended definition of concurrence for such (mixed) states\cite{woot}:
\begin{dex}[Concurrence: General definition]\label{gencon}
The
{\em concurrence} ${\cal C}$  of a  (mixed) two-qubit state $\rho$
is given by
\begin{equation}
{\cal C} = \max \left\{ \lambda _1-\lambda _2-\lambda _3-\lambda_4,0\right\},
\label{eq:c1}
\end{equation}
where the quantities $\lambda _i$ are the square roots of the eigenvalues of
the $4 \times 4$ matrix
\begin{equation}
\rho (\sigma _2\otimes \sigma _2)\rho^{*}(\sigma
_2\otimes \sigma_2)  \label{eq:c2}
\end{equation}
in descending order, where $\sigma_2=\left[\begin{array}{cc}0&-i\\i&0 \end{array} \right]$ .
\end{dex}
This applies whether   the density matrix  $\rho$ is
either pure or mixed; for pure states  it reduces to the form given in Definition \ref{concurrence}.
As already implied in the footnote given after Definition \ref{concurrence}, the entanglement of formation is a monotonic function of the concurrence
${\cal C}$, varying between a minimum of zero for ${\cal C}=0$,
and a maximum of 1 for ${\cal C}=1$.
\subsection{Decoherence of Bell State}
We now give an example of a standard dephasing process acting on a maximally entangled state.
\begin{example}\label{bcoh}
Consider the Bell state $v_B=1/\sqrt{2}[1,0,0,1]$.
The Liouville vector ${\bf r}$ corresponding to this is $1/2[1,0,0,1,0,0,0,0,0,0,0,0,1,0,0,1]$.
The action of the dephasing operator $L_{D0}$ of Eq.(\ref{LD0}) is given, as in Eq.(\ref{veceq}), by
 \begin{equation}\label{veceq0}
 \dot{\bf{r}} = L {\bf{r}} = (L_{D0}) {\bf{r}}
\end{equation}
which may be immediately integrated to give
\begin{equation}\label{}
{\bf r}(t)=1/2[1,0,0,{e^{-{\Gamma}_{{14}}t}},0,0,0,0,0,0,0,0,{e^{-{
\Gamma}_{{14}}t}},0,0,1]
\end{equation}
corresponding to the density matrix
\begin{equation}\label{}
 \rho(t)=1/2 \left[ \begin {array}{cccc} 1&0&0&{e^{-{\Gamma}_{{14}}t}}
\\\noalign{\medskip}0&0&0&0\\\noalign{\medskip}0&0&0&0
\\\noalign{\medskip}{e^{-{\Gamma}_{{14}}t}}&0&0&1
\end {array} \right].
\end{equation}
Note that this does not represent a pure state except at $t=0$.
The concurrence as defined in Definition \ref{gencon} evaluates to $\exp(-\Gamma_{14}t)$. (See Figure 2.)
\end{example}
\begin{figure}
\vspace{1cm}
\begin{center}\resizebox{8 cm}{!}
{\includegraphics{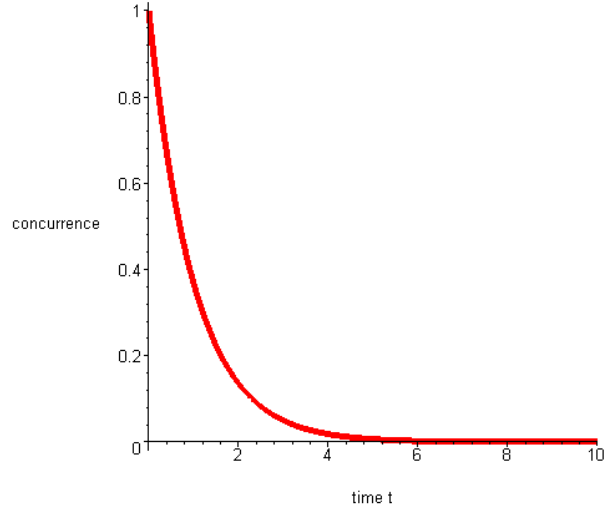}}
\caption{Concurrence {\it v} Time t (units of 1/Decoherence $\Gamma_{14}$)}
\end{center}
\label{conplot}
\end{figure}

The results of Example \ref{bcoh} are essentially unchanged in the presence of an additional  {\em free} hamiltonian, since this commutes with the dissipation super-operator $L_{D0}$, and indeed commutes with $L_D$ in general \cite{sol},  and only introduces a phase factor.
\subsection{Stable Bell state}
In general, entanglement will decay under the type of dissipative processes noted here.  However, as is clear from the last example, under special values of the decoherence parameters, entanglement will be preserved.  In the case of Example \ref{bcoh} when $\Gamma_{14} =0 (= \Gamma_{41})$ then the maximal entanglement of the  state does not decay. Of course, in general we are not able to specify the values of the dephasing $\Gamma$'s, but one may predict theoretically which types of state will remain invariant under the appropriate decoherence parameters. 

\end{document}